\documentclass[twocolumn, letterpaper,10pt]{IEEEtran}

\IEEEoverridecommandlockouts

\usepackage{rotating}
\usepackage{pdflscape}
\usepackage{graphicx}
\usepackage[english]{babel}
\usepackage{amssymb}
\usepackage{amsbsy}
\usepackage{verbatim}
\usepackage{cuted}
\usepackage{amsmath}
\usepackage{amsthm}
\usepackage{multirow}
\usepackage{multicol}
\usepackage{array}
\usepackage{cite}
\usepackage{float}

\usepackage[noend,ruled]{algorithm2e}

\usepackage{tikz-timing}

\usepackage{tikz}
\usepackage{pgfplots}

\newcommand{\fixme}[2]{\ifx&#2&{\color{red}#1}\else{\color{red}FIXME\{}#1{\color{red}\}}\footnote{{\color{red}#2}}\PackageWarning{Fixme}{#1: #2}\fi}

\floatstyle{plain}
\newfloat{twocolequfloat}{b}{zzz}
\floatname{twocolequfloat}{Equation}

\usepackage{ifpdf}
\ifpdf
\usepackage[draft,pdfborder={0 0 0}]{hyperref}
\fi

\title{A 9.52~dB NCG FEC scheme and 164 bits/cycle low-complexity product decoder architecture}

\author{Carlo Condo, Pascal Giard, \emph{Member, IEEE}, Fran\c{c}ois Leduc-Primeau, \emph{Member, IEEE}, \\Gabi Sarkis and Warren J. Gross, \emph{Senior Member, IEEE}\\
Department of Electrical and Computer Engineering, McGill University, Montreal, QC, Canada}

\begin{document}

\maketitle

\begin{abstract}
Powerful Forward Error Correction (FEC) schemes are used in optical communications to achieve bit-error rates below $10^{-15}$. These FECs follow one of two approaches: concatenation of simpler hard-decision codes or usage of inherently powerful soft-decision codes. The first approach yields lower Net Coding Gains (NCGs), but can usually work at higher code rates and have lower complexity decoders. In this work, we propose a novel FEC scheme based on a product code and a post-processing technique. It can achieve an NCG of 9.52~dB at a BER of $10^{-15}$ and 9.96~dB at a BER of $10^{-18}$, an error-correction performance that sits between that of current hard-decision and soft-decision FECs. A decoder architecture is designed, tested on FPGA and synthesized in 65 nm CMOS technology: its 164 bits/cycle worst-case information throughput can reach 100 Gb/s at the achieved frequency of 609~MHz. Its complexity is shown to be lower than that of hard-decision decoders in literature, and an order of magnitude lower than the estimated complexity of soft-decision decoders.
\end{abstract}

\section{Introduction}
\label{sec:intro}

Optical communication systems rely on extremely high-speed links that require high degrees of reliability. A Bit Error Rate (BER) lower than $10^{-15}$ and speeds of up to 100~Gb/s are required by the ITU-G.709 standard, a standard that defines the specifications for Optical Transport Networks (OTNs), while even higher speeds are foreseen in next generation standards. 
To achieve such low BER requirements, powerful Forward Error Correction (FEC) schemes must be employed. Recent approaches to high-performance, high-speed error-correction schemes follow one of two paths: concatenation of (often algebraic) hard-decision codes \cite{Lee_ISOCC10,Smith_JLT12,Jian_GLOBECOM13} or soft-decision, iterative decoding of inherently more powerful codes, first among all, Low-Density Parity-Check (LDPC) codes \cite{Gallager_TIT62}. The latter produced high-gain FEC schemes, that however must rely on complex decoding architectures \cite{Onohara_JSTQE10,Sugihara_OFC13,Huawei_SDFEC,Fujitsu_SDFEC}.


For example in \cite{Jian_GLOBECOM13}, Bose-Chaudhuri-Hocquenghem (BCH) codes \cite{Bose_IC60} are concatenated in a braided scheme and decoded with a hard decision algorithm. The FEC of \cite{Jian_GLOBECOM13} is reported to achieve 9.35 dB of Net Coding Gain (NCG) at a Bit Error Rate (BER) of $10^{-15}$ with a 7\% code overhead. While no decoder architecture is proposed, the estimated latency of the decoding scheme is of 1.15 million bits. With similar overhead, the FEC proposed in \cite{Lee_ISOCC10} uses different BCH codes in a quasi-product structure, achieving high throughput and a 9.19\,dB NCG at a high cost in area occupation. The BCH-based product code proposed in \cite{Wang_CL12}, with a code length of $98$~kbits bits and rate of 0.937, achieves a 9.4 NCG at BER=$10^{-15}$, without implementation details.
Staircase concatenation \cite{Smith_JLT12} has been recently proposed as an efficient and powerful FEC for 100\,Gb/s OTNs.

Soft-decision FECs are a relatively recent addition to the FEC world for optical communication. Few soft-decision FECs have been proposed, and no decoder implementations were found in the literature. 
In \cite{Onohara_JSTQE10} two FECs are proposed, a concatenated scheme using Reed-Solomon and LDPC codes, and a triple concatenation of an LDPC code with two algebraic codes. With a total overhead of 20.5\%, it was shown that an NCG of 10.8 dB could be achieved. BCH codes and spatially-coupled LDPC codes are used in \cite{Sugihara_OFC13}: a 12 dB NCG is estimated at a BER of $10^{-15}$, obtainable with a 25.5\% overhead. The FEC described in \cite{Miyata_OECC13} concatenates a soft-decision code with a product code, yielding 11 dB NCG at BER=$10^{-15}$ with a $20.5\%$ overhead and a code length of millions of bits.


We introduce in this paper a powerful FEC scheme relying on a product code \cite{Elias_IRE54} based on algebraic component codes, that thus belongs to the first category of FECs for optical communications. The proposed FEC can reach very low BER with a code rate comparable with recent OTN FEC solutions. A high-speed, low-complexity decoder architecture for the proposed FEC is designed, tested on a Field Programmable Gate Array (FPGA) and synthesized in 65 nm CMOS technology. We show that our decoder can reach a minimum 100\,Gb/s of information throughput at a frequency of 609~MHz, and has a gate count of approximately 1.15 million gates. It has a decoding latency of $319$~ns making it suitable for low-latency environments, like data centers.

The rest of this paper is organized as follows. Section~\ref{sec:FEC} describes the FEC scheme in details, its decoding process and its error-correction performance. In Section~\ref{sec:prodDec} the decoder hardware architecture is portrayed, while implementation and test results are given in Section \ref{sec:res}. Section~\ref{sec:mods} briefly discusses possible modifications to the decoder architecture along with their implications. Finally, Section~\ref{sec:conc} draws the conclusions.

\section{FEC Scheme}
\label{sec:FEC}


Product codes \cite{Elias_IRE54} are a class of error-correction codes constructed by encoding a matrix of information symbols row-wise with a row \emph{component} code, and subsequently column-wise using a column \emph{component} code. 
The twofold encoding acts as a parallel concatenation of the row and column component codes. The choice of the component code has a great impact not only on the error-correction performance of the product code, but on the speed and encoding/decoding complexity of the FEC scheme as well. 
BCH codes \cite{Bose_IC60} are a class of widely used algebraic codes, identified by the set of parameters $(n,k,t)$, where $n$ is the code length, $k$ the number of information bits, and $t$ the maximum number of errors that are guaranteed to be correctable. 
The standard BCH decoding algorithm relies on hard decision, and when $t=2$ (and to a lesser extent $t=3$), the general algorithm can undergo substantial simplifications~\cite{Gorenstein,Smith_JLT12} that reduce both latency and implementation complexity.
We thus consider BCH codes as a starting point for the construction of our FEC scheme.


	While it is not strictly necessary, we assume that the same BCH component code is used to encode both the rows and the columns of the information matrix.  We form a $k \times k$ matrix with the information bits. 
	Each row of the matrix is first encoded into a BCH code, resulting in a $k \times n$ matrix. Then, each of the $n$ columns are also encoded into a BCH code to form the $n \times n$ product-code codeword.
	Since the BCH component code is systematic, the product code is also systematic.
	Note that it is equivalent to first encode the columns of the information matrix, followed by the rows.
	We denote by $N=n^2$ the length of the resulting product code, and by $K=k^2$ the number of information bits in a codeword. The code rate of the product code is $K/N=k^2/n^2$.
	
	While a BCH code with $t=2$ guarantees a simple decoding process, a very long product code would be necessary to even get close to OTN's BER requirements. However, the error correction performance of the product code can be substantially improved at a small cost in code rate by using extended-BCH (eBCH) codes as component codes. An eBCH code of length $n$ is composed of a BCH code of length $n-1$ and of an additional parity bit, which increases the minimum distance of the code by 1. This increased distance can be used to reduce the probability of undetected failure of the component decoder, thereby reducing the number of new errors that are introduced by the component decoder and improving the performance of the product decoder.
	
	
		
	Since optical communications require a BER lower than $10^{-15}$, we must make sure that no error floor occurs at higher BER. The existence of an error floor is usually caused by particular error patterns that are difficult to impossible for the decoder to correct.
	A post-processing technique that can greatly enhance the error-correction performance of product codes based on polynomial component codes has been proposed in \cite{Condo_GlobalSip16}. The product code decoding is performed by alternatively decoding the rows and the columns of the received matrix: it is thus possible to identify rows and columns whose decoding has failed (see Section \ref{subsec:decalg} for more details). Based on this knowledge, the post-processing technique flips the bits at the intersection of failed rows and columns, greatly reducing the contribution of stall patterns to the error floor. This method is also applied in our proposed FEC scheme.

	\subsection{Decoding Algorithm} \label{subsec:decalg}

	As previously mentioned, the decoding of product codes can be performed by iteratively decoding the row and column component codes. Each iteration is divided into two half iterations, the first half decoding the rows and the second half the columns.
	Each row and column of the product code is decoded using the eBCH decoder described in Algorithm~\ref{alg:extendedbch}. The additional parity bit in the eBCH codeword is placed at position $n$. The $\mathrm{bch}(\cdot)$ function refers to the standard bounded distance BCH decoder, which returns a flag $\textsc{fail}$ indicating whether or not the decoder detected a failure, and a vector $e$ of length $n-1$ indicating the location of errors, if applicable. The notation $x_{i:j}$ with $i\leq j$ refers to a vector of length $j-i+1$ containing elements $i,i+1,\dots,j$ of the vector $x$. The operator $\oplus$ denotes modulo-2 addition. 
 
    
		\begin{algorithm}[tb] 
	   \SetKwInOut{Input}{input} \SetKwInOut{Output}{output}
	   \DontPrintSemicolon

	   \Input{Component codeword $r$}
	   \Output{Updated codeword $r'$}
	   
	   \Begin{
	   	$\textsc{fail}, e \gets \mathrm{bch}(r_{1:n-1})$ \;
		\If{$\textsc{fail}$}{
			$r' \gets r$ \tcp*[f]{decoding failure}\;
		}
		\Else{
			$d:= \sum_{i=1}^{n-1} e_i$ \;
        			$d_e:= \left(d + \sum_{i=1}^{n} r_i\right) \mod 2$ \;
        		
	        		\If{$d+d_e \leq t$}{
        				$r'_{1:n-1} \gets r_{1:n-1} \oplus e$ \;
        				$r'_{n} \gets r_{n} \oplus d_e$ \tcp*[f]{parity correction}\;
        			}
			\Else{
				$r' \gets r$ \tcp*[f]{decoding failure}\;
			}
		}
	   }
	   \caption{Decoding of eBCH codes}
	   \label{alg:extendedbch}
\end{algorithm}
	
	The BCH decoder can correct up to $t$ errors. If there are more than $t$ errors, the decoder could return another codeword, introducing an undetected failure. However, the parity extension allows detecting failures caused by the presence of $t+1$ errors. The eBCH decoder therefore declares a failure if either the BCH decoder detects a failure, or if $t+1$ errors are detected, i.e., if $d+d_e = t+1$.
		
	The post processing is applied after a predefined number of decoding iterations have been completed. Let us denote by $R$ ($C$) the set of row (column) indices for which the component decoder reported a decoding failure.
	If $0<|R|\le t+1$ and $0<|C|\le t+1$, we flip all the bits located at the intersection of a row in $R$ and of a column in $C$. Since this may introduce new bit errors, we then decode again all rows and columns whose bits were flipped.

	When $t=2$, the decoding of the BCH part of eBCH component codes can be substantially simplified by using the Peterson-Gorenstein-Zierler algorithm~\cite{Gorenstein}. As will be shown in Section~\ref{subsec:perf}, codes with $t=2$ can achieve very good error-correction performance even at moderately high rates: at the same time, the decoder architecture benefits from reduced complexity and latency. Thus, the $\mathrm{bch}(\cdot)$ function relies on the specialized algorithm, that differs from standard BCH decoding algorithms \cite{Bose_IC60} in that syndrome values are used directly to find the roots of the error-locator polynomial. Only two syndromes need to be calculated:
    \begin{align}
      S_1 &= \sum_{i = 0} ^ {n - 1} r_i \alpha^i, \label{eq:s1}\\
      S_3 &= \sum_{i = 0} ^ {n - 1} r_i \alpha^{3i}; \label{eq:s3}
    \end{align}
    where $r$ is the input to the decoder and $\alpha$ the primitive element of the BCH Galois Field (GF).
    Based on the values of $S_1$ and $S_3$, different cases arise:
    \begin{itemize}
    \item $S_1 = 0$ and $S_3 = 0$: no errors were detected.
    \item $S_1 \neq 0$ and $S_1^3 + S_3 = 0$: one error located at $\log_\alpha S_1$ was detected.
    \item $S_1 = 0$ and $S_1^3 + S_3 \neq 0$: more than two errors occurred and the decoder declares failure.
    \item $S_1 \neq 0$ and $S_1^3 + S_3 \neq 0$: two or more errors occurred. In this case, the decoder attempts to find the roots ($\rho_1$ and $\rho_2$) of
      \begin{equation}
        \label{eq:special-k}
        x^2 + x + \frac{S_1^3 + S_3}{S_1^3} = 0.
      \end{equation}
      Decoding failure is declared if no roots were found. Otherwise, the decoder detects two errors located at $\log_\alpha S_1 \rho_1$ and $\log_\alpha S_1 \rho_2$.
    \end{itemize}

\subsection{Code Selection and Error-Correction Performance}
\label{subsec:perf}

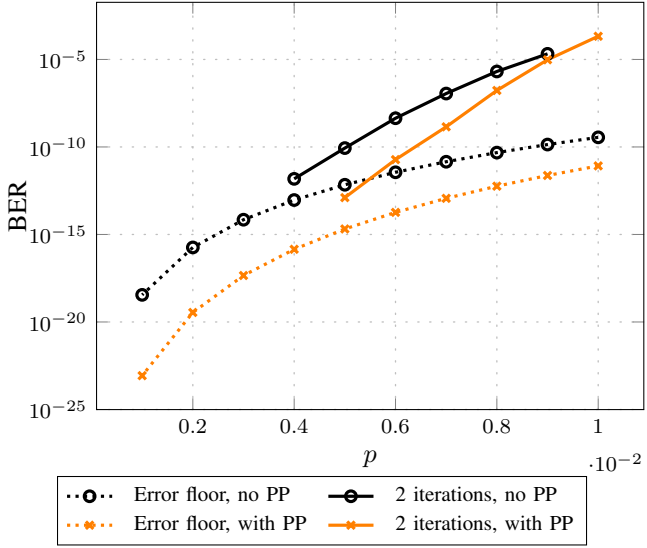
\begin{figure}[tbp]
  \begin{center}




\begin{tikzpicture}

  \pgfplotsset{
    grid style = {
      dash pattern = on 0.05mm off 1mm,
      line cap = round,
      line width = 0.5pt,
      loosely dotted
    },
    label style = {font=\fontsize{10pt}{7.2}\selectfont},
    tick label style = {font=\fontsize{8pt}{7.2}\selectfont},
  }

  \begin{semilogyaxis}[%
    xlabel=$p$,%
    ylabel=BER, ylabel style={yshift=-0.5em},%
    every axis x label/.style={at={(ticklabel cs:0.5)},anchor=near ticklabel},
    width=\columnwidth, height=7cm, grid=major,%
    legend style={
      anchor={center},
      cells={anchor=west},
      column sep=2mm,
      font=\fontsize{8pt}{7.2}\selectfont,
      mark options=solid,
    },
    mark options=solid,
    legend to name=bch-bound-legend,
    legend columns=2,
    grid=major,
    xminorgrids=true,
    yminorgrids=true,
    ]

    \addplot[color=black, dotted, mark=o, very thick] table[x=p,y=BER]{Figures/data/tpc_195-178-2_errorfloor_nopp.txt};
    \addlegendentry{Error floor, no PP}

    \addplot[color=black, very thick, mark=o] table[x=p,y=BER]{Figures/data/tpc_195-178-2_2iter_nopp.txt};
    \addlegendentry{2 iterations, no PP}

    \addplot[color=orange, dotted, mark=x, very thick] table[x=p,y=BER]{Figures/data/tpc_195-178-2_errorfloor_pp1.txt};
    \addlegendentry{Error floor, with PP}

    \addplot[color=orange,mark=x, very thick] table[x=p,y=BER]{Figures/data/tpc_195-178-2_2iter_pp1_cd3f357.txt};
    \addlegendentry{2 iterations, with PP}
  \end{semilogyaxis}
\end{tikzpicture}
\\
\ref{bch-bound-legend}

    \caption{Error floor estimation and BER curves for an extended BCH-based (195,178,2)$^2$ product code over a BSC.}
    \label{fig:noPPvsPP}
  \end{center}
\end{figure}

Depending on the requirements, the proposed FEC scheme can employ different eBCH component codes. We have evaluated the effect of different code parameters on both the simulated BER and the estimated error floor.
Existing FEC schemes for optical communications vary in code length, rate and decoding complexity. The recent trends towards soft-decision decoding led to high NCGs, with code overheads reaching 20\% and large estimated decoder area occupations \cite{Onohara_OFC10,Huawei_SDFEC,Fujitsu_SDFEC}.
An overhead of 20\% translates into a code rate of approximately 0.833. For our proposed FEC, using the extended-BCH (256,239,2) code as a component code, the resulting product code has a rate of 0.878. We can thus consider shortening the code by $\ell$ bits, leading to a product code of rate $\frac{(k-\ell)^2}{(n-\ell)^2}$. For rates greater than 0.833, with $n=256$ and $k=239$, the shortening can use any $\ell \leq 61$. Using $\ell=61$, the resulting product code has a length of $(256-61)^2=38,025$ bits.



Fig.~\ref{fig:noPPvsPP} plots the BER for the $(195, 178)^2$ product code, along with the error floor, estimated as in \cite{Condo_GlobalSip16}, with and without the use of post processing. The reported error floor represents the contribution of minimal stall patterns to the error rate.
Simulations have been performed on a binary symmetric channel (BSC), and $p$ represents the input error probability.
It can be seen that both the error floor and BER of the considered product code are substantially reduced by post processing. As $p$ decreases, the BER approaches the estimated error floor, which has been shown to be a tight lower bound on the BER for this code \cite{Condo_GlobalSip16}. Table \ref{tab:NCG} reports the NCG values achieved by the proposed FEC at different values of $p$: at the commonly considered BER of $10^{-15}$, the bound shown by our FEC has an NCG of 9.52~dB that grows up to 9.95~dB at a BER of $10^{-18}$. As shown in \cite{Condo_GlobalSip16}, the BER curve reaches the bound earlier than BER=$10^{-13}$ when four decoding iterations are performed: the trend shown in Fig. \ref{fig:noPPvsPP} lets us assume that the bound will be reached at around BER=$10^{-15}$ or slightly lower when two decoding iterations are considered.

\begin{figure}[t]
  \begin{center}
    \definecolor{darkgreen}{RGB}{0, 128, 0}


\begin{tikzpicture}
  \pgfplotsset{
    grid style = {
      dash pattern = on 0.05mm off 1mm,
      line cap = round,
      line width = 0.5pt,
      loosely dotted
    },
    label style = {font=\fontsize{10pt}{7.2}\selectfont},
    tick label style = {font=\fontsize{8pt}{7.2}\selectfont}
  }

  \begin{semilogyaxis}
    [
    width=\columnwidth, height=7cm,
    line cap=round,
    every axis x label/.style={at={(ticklabel cs:0.5)},anchor=near ticklabel},
    xmin=1e-3,
    xmax=1.5e-2,
    xlabel={$p$},
    minor x tick num={1},
    ylabel=BER, ylabel style={yshift=-0.5em},%
    legend columns = 2,
    legend style={
      anchor={center},
      cells={anchor=west},
      column sep= 2mm,
      font=\fontsize{8pt}{7.2}\selectfont,
      mark options=solid,
    },
    mark options=solid,
    legend to name=ber-exp-legend,
    grid=major,
    xminorgrids=true,
    yminorgrids=true,
    grid style=loosely dotted,
    ]

    \addplot[color=black,very thick, mark=x] table[x=p,y=BER] {Figures/data/195.178.i2.no-pp.o7777};
    \addlegendentry{$(195, 178)^2$ 2 it., no PP}

    \addplot[color=black,very thick, dotted, mark=x] table[x=p,y=BER] {Figures/data/tpc_195-178-2_2iter_pp1_cd3f357.txt};
    \addlegendentry{$(195, 178)^2$ 2 it., with PP}

    \addplot[color=purple,very thick, mark=triangle ] table[x=p,y=BER] {Figures/data/tpc_195-178-2_4iter_cd3f357.txt};
    \addlegendentry{$(195, 178)^2$ 4 it., no PP}

    \addplot[color=purple,very thick, dotted, mark=triangle] table[x=p,y=BER] {Figures/data/tpc_195-178-2_4iter_pp1_cd3f357.txt};
    \addlegendentry{$(195, 178)^2$ 4 it., with PP}

    \addplot[color=blue,very thick, mark=o] table[x=p,y=BER] {Figures/data/219.200.i2.no-pp.o7786};
    \addlegendentry{$(219, 200)^2$ 2 it., no PP}

    \addplot[color=blue,very thick, dotted, mark=o] table[x=p,y=BER] {Figures/data/219.200.i2.pp.o7785};
    \addlegendentry{$(219, 200)^2$ 2 it., with PP}

    \addplot[color=red,very thick, mark=+] table[x=p,y=BER] {Figures/data/tpc_321-293-3_2iter.txt};
    \addlegendentry{$(321, 293)^2$ 2 it., no PP}

    \addplot[color=brown,very thick, mark=square] table[x=p,y=BER] {Figures/data/tpc_321-293-3_4iter.txt};
    \addlegendentry{$(321, 293)^2$ 4 it., no PP}

  \end{semilogyaxis}

\end{tikzpicture}
\\
\ref{ber-exp-legend}

    \caption{Code parameter variation effect on BER curves for extended BCH-based product codes over a BSC, with a fixed 20\% overhead.}
    \label{fig:BERexp}
  \end{center}
\end{figure}
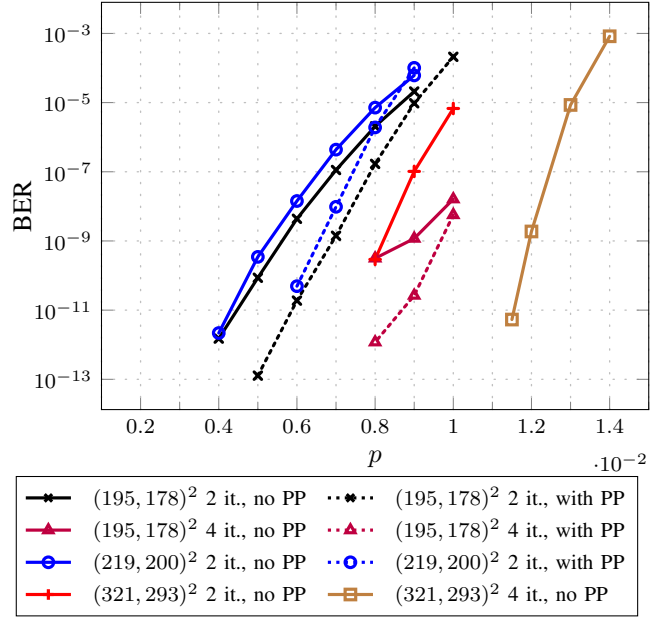

Fig.~\ref{fig:BERexp} shows how the error-correction performance changes as the code rate is kept constant, while $n$, $t$, the number of iterations and the application of post processing are varied. The BER of the $(195, 178)^2$ product code is shown for two and four decoding iterations, with and without the application of post processing. Increasing the number of iterations results in a substantial improvement at higher $p$ values. However, the main contribution to the error floor comes from error patterns that the decoder cannot correct, regardless of the number of iterations. Consequently, as $p$ decreases, the two and four iteration curves converge. This trend can be observed with and without post processing.

The $(219,200)^2$ product code uses a component code shortened from the $(512,~493,~2)$ eBCH code. It is 26\% longer than the $(195, 178)^2$ code. The large amount of applied shortening slows the convergence speed of this code: its curve slope is bound to outperform the $(195, 178)^2$ curve at around BER=$10^{-12}$. Thus, a larger number of iterations is necessary to fully exploit this code at higher $p$, decreasing the achievable throughput. Moreover, the decoder architecture would need a significant amount of additional memory, and the trade-off between logic and latency would be less advantageous.
Two and four iterations BER curves for a $(321, 293)^2$ product code are plotted as well: it is the smallest product code with $t = 3$ and the same rate as the $(195, 178)^2$ product code. It is 171\% longer than the $(195, 178)^2$ code. Its error-correction performance is better than the other codes shown in Fig.~\ref{fig:BERexp}. However, a decoder architecture targeting this code would be significantly more complex. In fact, aside from the use of $t = 3$ requiring slightly higher decoding and hardware complexity than $t = 2$, the longer code would substantially increase gate count and decoding latency.

\begin{table}
	\centering
	\caption{Net Coding Gain values for the proposed FEC.}
	\label{tab:NCG}
  \setlength{\extrarowheight}{1.8pt}
  \begin{tabular}{|cc|c|}
    \hline
    $p$ & BER  & NCG \\ 
        & &   [dB] \\ 
    \hline\hline
    $7\cdot10^{-3}$ &  $10^{-9}$   & 7.7507 \\ 
    $5\cdot10^{-3}$ &  $10^{-13}$  & 9.1061 \\ 
    $4\cdot10^{-3}$ &  $10^{-15}$    &  9.5260\\ 
    $2.7\cdot10^{-3}$ &  $10^{-18}$   & 9.9596 \\ 
    \hline
  \end{tabular}
\end{table}

\section{Product Decoder Architecture}
\label{sec:prodDec}

\begin{figure}[t!]
\centering
	\includegraphics[scale=0.4]{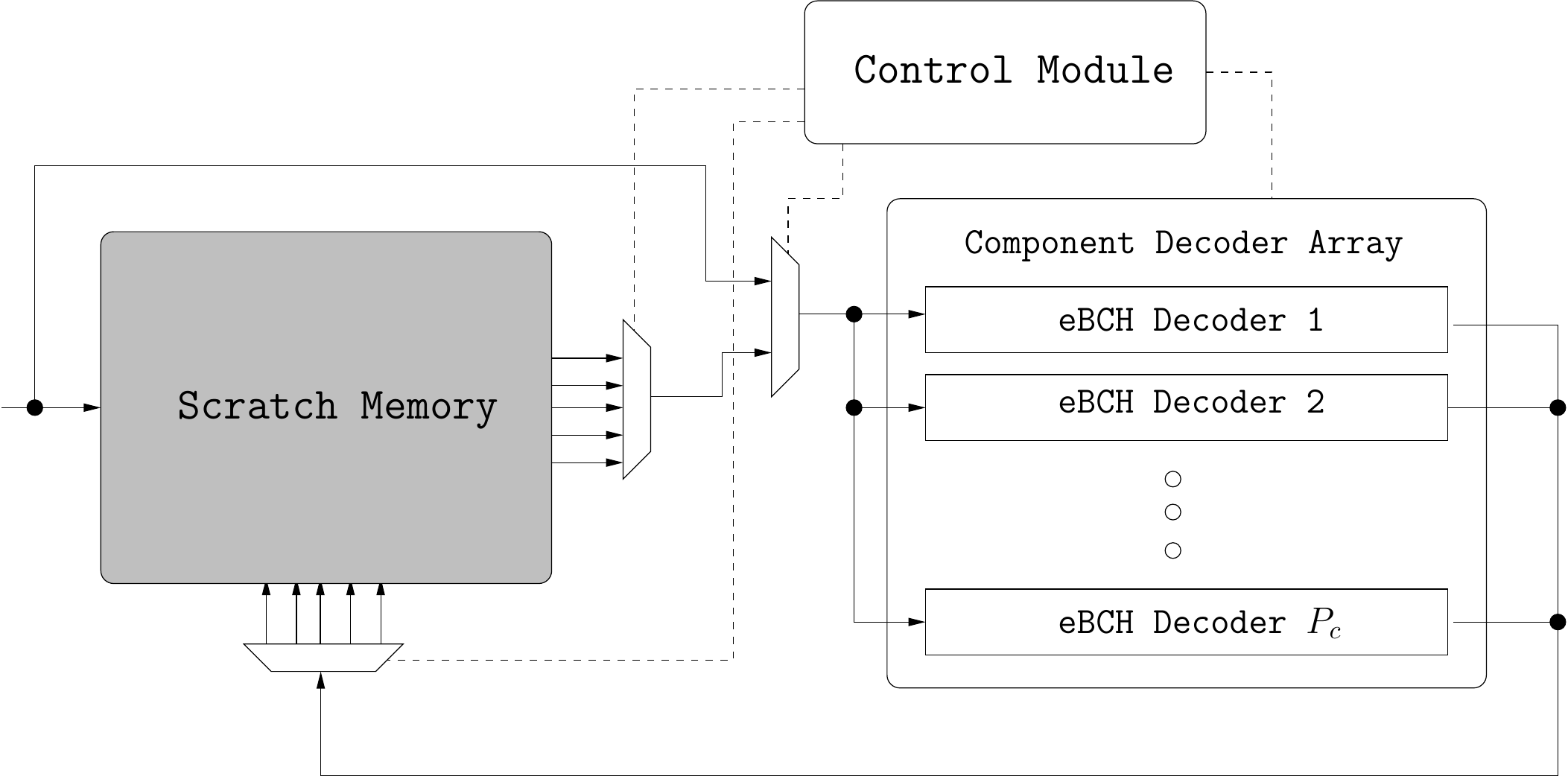}
	\caption{Product decoder Architecture.}
	\label{fig:PD}
\end{figure}

The overall structure of the product decoder is portrayed in Fig.~\ref{fig:PD}. The product code is stored in a $n \times n$ register matrix acting as a scratch memory. The proposed architecture is sized on the considered $(195, 178)$ component code: Section~\ref{sec:mods} discusses the necessary modifications in case the code is changed.
An array of $P_c$ component decoders decodes as many product code rows (columns) in parallel. Inputs and outputs of each eBCH decoder are connected to $\frac{n}{P_c}$ rows and $\frac{n}{P_c}$ columns of the scratch memory. The outputs of the component decoders flip the bits in the scratch memory that are identified as incorrect: they are ANDed with a valid signal coming from the control module, while the inputs to the component decoders are multiplexed, scanning the rows and the columns in order. The control of the decoder architecture can be greatly simplified in case $P_c$ is an exact divisor of $n=195$: the proposed architecture has consequently been sized for $P_c=13$, a choice offering a good tradeoff between achievable throughput and hardware complexity.

Product codewords are loaded from an external input buffer into the scratch memory, through a bus as wide as $P_l$ eBCH codewords ($P_l\times n$ bits). This bus is also connected to the component decoder array, allowing the first half iteration to be performed in parallel to the codeword loading. Each register of the scratch memory is preceded by an XOR gate, that allows the bit-flipping signals coming from the component decoders to correct errors. The proposed architecture has been sized assuming $P_l=2$.

The scratch memory features two $n$-bit failure registers that keep track of which rows and columns have suffered a decoding failure during the last half iteration in which they were involved.


In Section \ref{sec:eBCHarch} to \ref{subsec:PPit}, we detail the product-decoder architecture and its operation. In particular, we detail the eBCH component decoder, and then divide the decoding process into three conceptual functions: the loading of the product codeword and first half iteration, the standard iterations and the post-processing iteration.

\subsection{Extended-BCH Decoder Architecture}
\label{sec:eBCHarch}

\begin{figure*}[ht]
	\begin{center}
	\includegraphics[width=\textwidth]{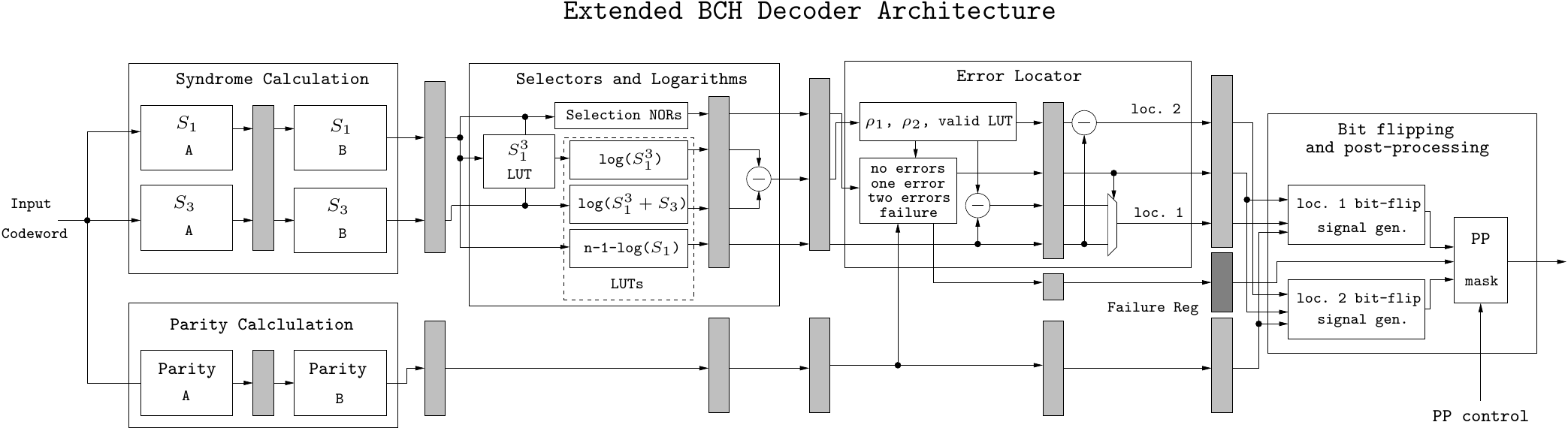}
	\caption{eBCH decoder architecture.}
	\label{fig:eBCHdec}
	\end{center}
\end{figure*}

In this section, we describe the designed eBCH decoder architecture, whose functional scheme is portrayed in Fig. \ref{fig:eBCHdec}. 

Five main blocks can be identified: the \texttt{syndrome calculation} module, that works in parallel to the \texttt{parity calculation} module, the \texttt{selectors and logarithms} module, the \texttt{error locator} module and the \texttt{bit-flipping and post-processing} module. Light gray blocks represent pipeline stages, while the darker gray block is the failure register (described in details in Section~\ref{subsubsec:FReg}).

\subsubsection{Syndrome Calculation Module}
\label{subsec:syn}

The \texttt{syndrome calculation} module performs \eqref{eq:s1} and \eqref{eq:s3} in parallel on the BCH codeword. All $\alpha^i$ and $\alpha^{3i}$ are precomputed and stored as static 8-bit values. Since $r_i$ is a single bit, each multiplication in $r_i\alpha^i$ and $r_i\alpha^{3i}$ requires 8 AND gates. Summations within GF(8) are equivalent to the XOR operation, so each sum in \eqref{eq:s1} and \eqref{eq:s3} requires 8 XOR gates. The XOR tree required to perform them all is split between the fourth and fifth stages to shorten the critical path.

\subsubsection{Parity Calculation Module}
\label{subsec:par}

The \texttt{parity calculation} module performs $\sum_{i=1}^{n} r_i$, that requires XORing all $n$ codeword bits. As this module works in parallel to the \texttt{syndrome calculation} module, and its structure is similar, an internal pipeline stage splits the XOR tree between the fourth and fifth stages as well. 

\subsubsection{Selectors and Logarithms Module}
\label{subsec:SAL}

This module performs partial calculations and logarithmic domain conversions that are needed by the \texttt{error locator} module to identify errors. 
Four 8-bit-wide Lookup Tables (LUTs) are needed to calculate the following quantities:
\begin{itemize}
\item $S_1^3$, with input $S_1$;
\item $\log(S_1^3)$, with input $S_1^3$;
\item $n-1-\log(S_1)$, with input $S_1$;
\item $\log(S_1^3+S_3)$, with input $S_1^3+S_3$.
\end{itemize}
Since both $\log(S_1^3)$ and $\log(S_1^3+S_3)$ perform the same operation with different inputs, they are merged into a single LUT. 

The summation required by $S_1^3+S_3$ is performed within GF(8), requiring 8 XOR gates. An 8-bit adder is instead required to perform $\log(S_1^3+S_3) - \log(S^3)$: switching to logarithmic domain allows to avoid a division, but sums are not constrained to GF(8) anymore, and cannot be implemented with an XOR operation.
The \texttt{Selection NORs} block in Fig. \ref{fig:eBCHdec} evaluates the following signals, each of which can be calculated with an 8-input NOR gate:
\begin{itemize}
\item $S_1^z = 1$ if $S_1=0$;
\item $S_3^z = 1$ if $S_3=0$;
\item $(S_1^3+S_3)^z=1$ if $S_1^3+S_3=0$.
\end{itemize}
These three signals are passed to the \texttt{error locator} module, along with $n-1-\log(S_1)$ and $\log(S_1^3+S_3) - \log(S^3)$. 

To reduce the system's critical path, an internal pipeline is present in this module. All LUTs are placed before the pipeline stage, along with most calculations, except $\log(S_1^3+S_3) - \log(S^3)$, that is performed after the registers.

\subsubsection{Error Locator Module}
\label{subsec:ELO}

The \texttt{error locator} module is tasked with the solution to \eqref{eq:special-k} and the unequivocal identification of the status of the eBCH decoding process (no errors, one error, two errors, failure).
A 17-bit-wide LUT stores the values of $\log(\rho_1)$ and $\log(\rho_2)$, i.e. the logarithm of the roots of \eqref{eq:special-k}, along with a validity flag to signal if the roots exist or not. The LUT is addressed through $\log(S_1^3+S_3) - \log(S^3)$. Two 8-bit adders compute $(n-1-\log(S_1)) - \log(\rho_1)$ and $(n-1-\log(S_1)) - \log(\rho_2)$,  the error locations in case the decoder detects two errors. The error location in case of a single error is $n-1-\log(S_1)$.

The decoder status is determined on the basis of the signals computed in the \texttt{selectors and logarithms} module, the parity check result, and the validity of the computed roots, through the following set of boolean equations:
\begin{equation*}
{\rm \bf NoErrors}: S_1^z \land S_3^z
\end{equation*}
\begin{equation*}
{\rm Fail_1}: S_1^z \land \overline{S_3^z} \land \overline{(S_1^3+S_3)^z}
\end{equation*}
\begin{equation*}
{\rm 1Error_1}: (S_1^3+S_3)^z \land \overline{S_1^z}
\end{equation*}
\begin{equation*}
{\rm 2Errors_1}: \big(S_1^z \land \overline{S_3^z} \land (S_1^3+S_3)^z \big) \lor \big(\overline{S_1^z} \land \overline{(S_1^3+S_3)^z} \big) 
\end{equation*}
\begin{equation*}
{\rm Fail_2}: {\rm 2Errors_1} \land \big((\sum_{i=1}^{n} r_i \land {\rm ValidRoots})  \lor \overline{{\rm ValidRoots}}\big) 
\end{equation*}
\begin{equation*}
{\rm Fail_3}: {\rm 2Errors_1}~ \land ~\big( {\rm ErrorLoc_1}>n-1 ~\lor~  {\rm ErrorLoc_2}>n-1    \big) 
\end{equation*}
\begin{equation*}
{\rm Fail_4}: {\rm 1Error_1} \land \big( {\rm ErrorLoc_1}>n-1 \big) 
\end{equation*}
\begin{equation*}
{\rm \bf Failure}: {\rm Fail_1} \lor {\rm Fail_2} \lor {\rm Fail_3} \lor {\rm Fail_4}
\end{equation*}
\begin{equation*}
{\rm \bf OneError}:{\rm 1Error_1} \land \overline{\rm Failure}
\end{equation*}
\begin{equation*}
{\rm \bf TwoErrors}:{\rm 2Errors_1}\land \overline{\rm Failure}
\end{equation*}
The four boldfaced signals are in mutual exclusion and are passed to the \texttt{bit-flipping and post-processing} module along with the two error locations. {\bf OneError} is used to select between the two possible error locations $(n-1-\log(S_1)) - \log(\rho_1)$ and  $(n-1-\log(S_1))$, and {\bf Failure} is stored in one of the two $n+1$-bit failure registers of the product code decoder, that track eBCH decoding failures among rows and columns.

As with the \texttt{selectors and logarithms} module, an internal pipeline stage reduces the system critical path. The validity of the roots, the second error location and the first four boolean equations are evaluated before the pipeline, while the other boolean equations and selection of the first error location are performed after the registers.

\subsubsection{Bit-Flipping and Post-Processing Module}
\label{subsec:BFPP}

According to the provided error locations, this module selects the appropriate signals to correct the errors by flipping bits. The bit-flipping signals are combined and masked following the decoder status and post processing.
Each error location is converted in a bit-flipping signal of $n$ bits, one-hot encoded, and masked according to the status of the decoder:
\begin{itemize}
\item No errors or failure: both bit-flipping signals are nulled through AND gates;
\item One error: the second error location is nulled through AND gates;
\end{itemize}
The additional parity bit-flipping signal is determined according to Alg. \ref{alg:extendedbch}. 

A post-processing activation signal is received as an input from the product-code decoder \texttt{control} module: it is activated in case $0<|R|< t+1$, and the eBCH decoder is currently performing the last decoding iteration on a column of the codeword matrix. Thus, if the status of the decoder is failure and post processing is active, the content of the row-failure register is substituted to the bit-flipping signal. If at the end of the product-decoder iteration $0<|C|\le t+1$, then a last iteration on the rows and columns in $R$ and $C$ is issued, otherwise decoding is declared unsuccessful.

\subsection{Codeword Loading and First Half Iteration}
\label{subsec:first}


The first half iteration can be run in parallel with the loading of the product codeword in the scratch memory. 
At the first clock rising edge after a reset, the loading of the product codeword and the first half iteration begins.


\begin{figure}[t!]
\centering
	\includegraphics[scale=0.63]{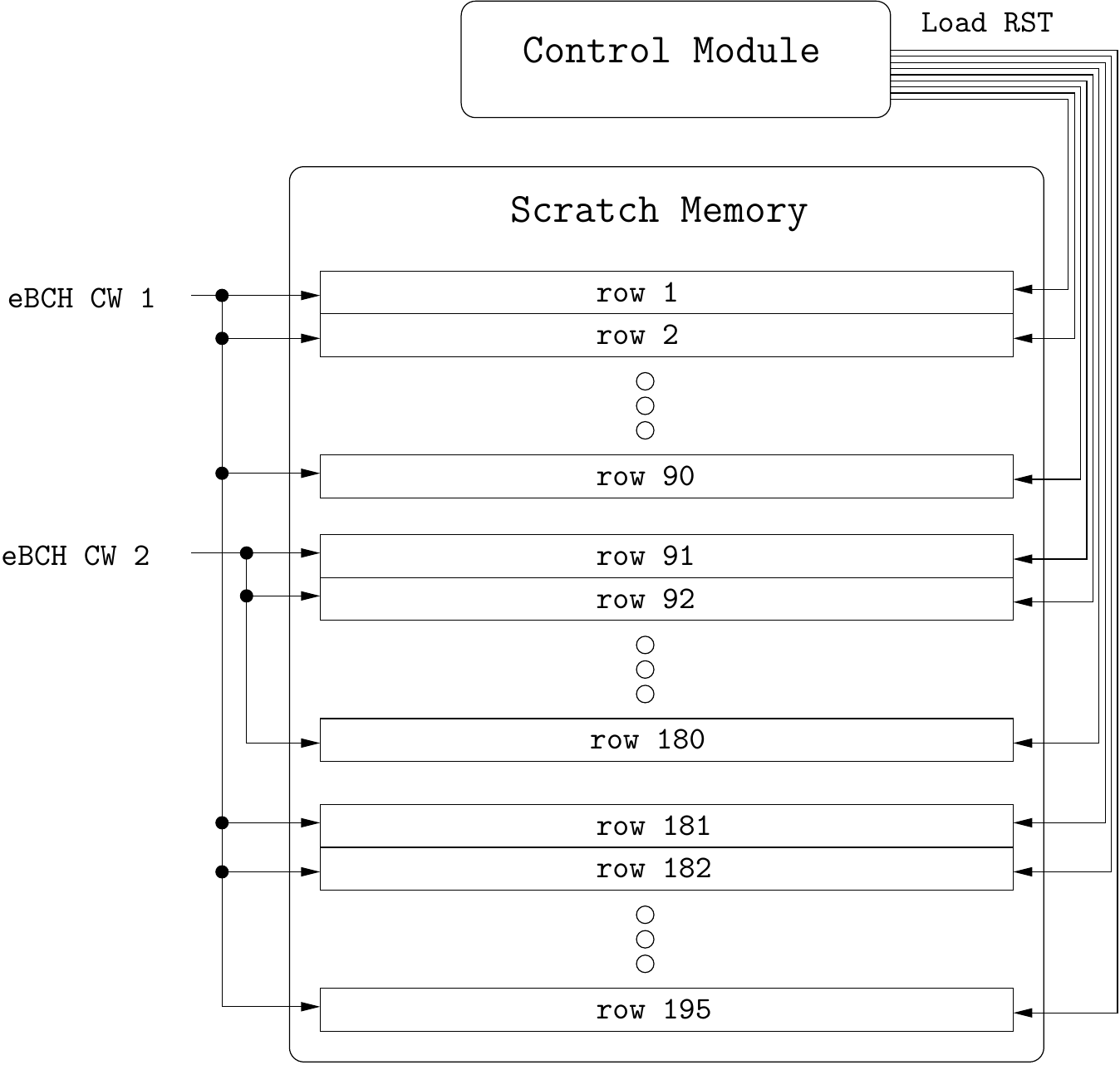}
	\caption{Product codeword loading.}
	\label{fig:load}
\end{figure}

The loading of the scratch memory is performed row wise, and is depicted in Fig.~\ref{fig:load}.
At each clock cycle, the \texttt{control} module issues up to two reset signals to the scratch memory. When a row is reset, its value is available at the decoder output for one clock cycle, while it is substituted with that of \texttt{eBCH CW 1} or \texttt{2}, depending on the row.

\begin{itemize}
 \item {\bf Clock cycle $1 \rightarrow 90$}: \texttt{eBCH CW 1} loaded in scratch memory rows $1 \rightarrow 90$, \texttt{eBCH CW 2} loaded in scratch memory rows $91 \rightarrow 180$. Scratch memory rows $1 \rightarrow 90$ output through \texttt{Output eBCH CW 1}, scratch-memory rows $91 \rightarrow 180$ output through \texttt{Output eBCH CW 2}.
 \item {\bf Clock cycle $91 \rightarrow 105$}: \texttt{eBCH CW 1} loaded in scratch memory rows $181 \rightarrow 195$. Scratch memory rows $181 \rightarrow 195$ output through \texttt{Output eBCH CW 1}. These 15 clock cycles could be reduced to 8 if both \texttt{eBCH CW 1} and \texttt{2} were used concurrently: however, all the rows $181 \rightarrow 195$ are connected to the same component decoder, thus 15 clock cycles will be required to use them as inputs anyway.
\end{itemize}

\begin{figure}[t!]
\centering
	\includegraphics[scale=0.55]{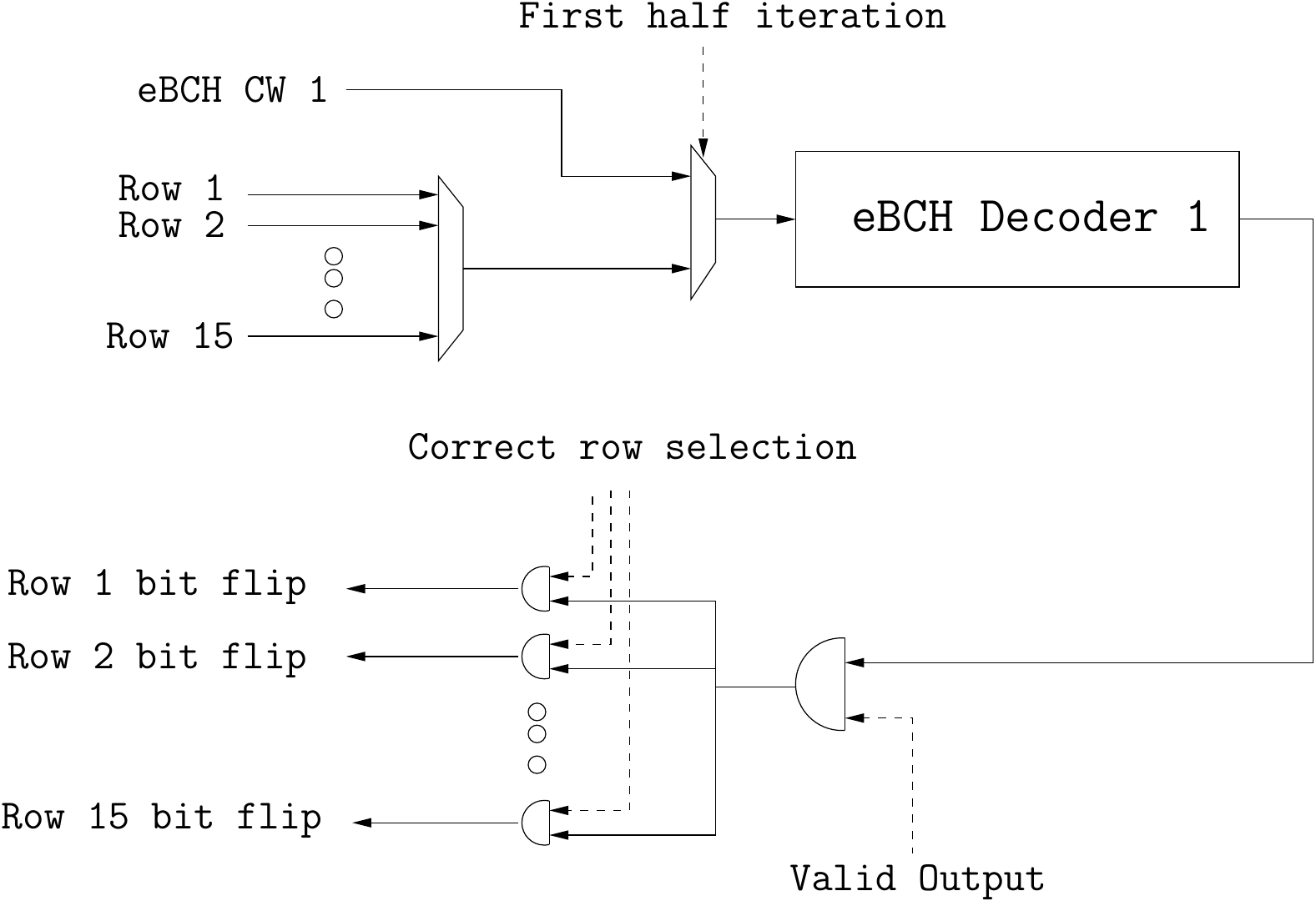}
	\caption{Input and output selection and validation for \texttt{eBCH decoder 1} during the first half iteration.}
	\label{fig:valid}
\end{figure}

During the first half iteration, the input of each component decoder is not one of the 15 rows of the scratch memory to which it is connected, but either \texttt{eBCH CW 1} or \texttt{2}, depending on the decoder.
In this way, the codewords currently being loaded in the scratch memory can bypass the loading itself, and directly be decoded. 
Fig.~\ref{fig:valid} shows the input multiplexing and output validation for the first component decoder in the array.
The multiplexing of inputs is static and does not change for the whole first half iteration, so that component decoder inputs are as follows:
\begin{itemize}
\item {\bf Clock cycle $1 \rightarrow 105$}: \texttt{eBCH CW 1} input to \texttt{eBCH $1 \rightarrow 6$} and {eBCH $13$}, \texttt{eBCH CW 2} input to \texttt{eBCH $7 \rightarrow 12$}.
\end{itemize}
On the other hand, even if all component decoders have received an input, their outputs must be enabled only for the correct scratch memory rows. Considering that the length of the pipeline within component decoders is that of 6 delay elements, the \texttt{Valid Output} signals issued by the control module follow this pattern:
\begin{itemize}
 \item {\bf Clock cycle $6 + 1 \rightarrow 15$}: \texttt{eBCH decoder 1} and \texttt{eBCH decoder 7} have valid outputs.
\item {\bf Clock cycle $6 + 16 \rightarrow 30$}: \texttt{eBCH decoder 2} and \texttt{eBCH decoder 8} have valid outputs.
\item {\bf Clock cycle $6 + 31 \rightarrow 45$}: \texttt{eBCH decoder 3} and \texttt{eBCH decoder 9} have valid outputs.  
\item {\bf Clock cycle $6 + 46 \rightarrow 60$}: \texttt{eBCH decoder 4} and \texttt{eBCH decoder 10} have valid outputs.
\item {\bf Clock cycle $6 + 61 \rightarrow 75$}: \texttt{eBCH decoder 5} and \texttt{eBCH decoder 11} have valid outputs.
\item {\bf Clock cycle $6 + 76 \rightarrow 90$}: \texttt{eBCH decoder 6} and \texttt{eBCH decoder 12} have valid outputs.
\item {\bf Clock cycle $6 + 91 \rightarrow 105$}: \texttt{eBCH decoder 13} has valid output.
\end{itemize}

The validated bit-flipping signal is itself zeroed for all the rows connected to the component decoder except for the correct one (see \texttt{Correct row selection} signals in Fig.~\ref{fig:valid}).
The component-decoder internal pipeline ensures that the loading of a codeword has been completed before the component decoder tries to correct it.

\subsection{Standard Iterations}
\label{subsec:stdit}

What we defined as standard iterations are the second, third and fourth half iterations. The second and fourth half iterations decode the columns of the product code, while the third decodes the rows. During these half iterations, all 13 component decoders work in parallel. Thus, each of these lasts $[(195/13)=15] + 6$ clock cycles, where 6 is the length of the component decoder pipeline.

The \texttt{currRowIn} signal is issued by the \texttt{control} module and scans the rows (columns) connected to each component decoder from 1 to 15, one per clock cycle, so that the input of each component decoder is the scratch memory row (column) identified by Eq. \eqref{eq:conn}:
\begin{equation}
 \label{eq:conn}
 {\rm Input~row~(column)}=({\rm n_{eBCH}}-1)\times15 + {\rm \texttt{currRowIn}}
\end{equation}
where ${\rm n_{eBCH}}$ is the number assigned to a component decoder within the component decoder array.

At the start of each half iteration, all component-decoder outputs are invalid, and are made valid simultaneously when the input data has reached the end of their internal pipeline. The selection of the correct row (column) for the output (see Fig.~\ref{fig:valid}) is made according to the \texttt{currRowOut} signal, that is the pipelined version of \texttt{currRowIn}.

\subsubsection{Failure Registers}
\label{subsubsec:FReg}
As mentioned before, the row- and column-failure registers are two $195$-bit registers that are used to track which rows and columns decoding has failed. 
The row- (column-) failure register is updated during all half iterations that decode scratch memory rows (columns). They are reset at the start of a corresponding half iteration, and updated with the value of the {\bf Failure} signal coming from all component decoders according to the value of \texttt{currRowOut}. 

Failure registers are used in different stages of the decoding process:
\begin{itemize}
 \item After the last half iteration, that is always a column half iteration, the column-failure register holds the most up-to-date information about the product-code decoding status. Consequently, the outcome of the decoding of the product codeword can be determined by ORing all the bits in the column-failure register: if the result is 1, at least a column has failed, and general decoding failure is declared. On the contrary, a success flag is raised if all bits in the failure register are zero.
 \item The row-failure register is used  at the beginning of the fourth half iteration to determine if post processing should be applied: details are given in Section \ref{subsubsec:PPA} below.
 \item The content of both registers is used to determine if the post-processing iteration would be useful or not. If both registers identify between one and three failures, then the post processing has been successfully applied and the post-processing iteration should be run. More details are provided in the following Sections \ref{subsubsec:PPA} and \ref{subsec:PPit}.
 \end{itemize}

\subsubsection{Post-Processing Application}
\label{subsubsec:PPA}
The idea behind post processing is that if the number of failed rows and columns is between one and three, some stalling patterns can be circumvented by flipping the bits at the intersection of failed rows and columns. Afterwards, the decoding of the previously failed rows and columns is attempted again.

The same result can be obtained in hardware using a slightly different schedule:
\begin{enumerate}
 \item At the end of the third half iteration, the row-failure register has a 1 in every position corresponding to a failed row.
 \item During the fourth half iteration, every time a column decoding fails, the column-failure register is updated. 
 In case of failure, the bit-flipping signal coming from the component decoder is the all-zero signal, i.e. no bits are flipped. However, if the number of ones in the row-failure register is between one and three, the bit-flipping signal is substituted with the content of the row-failure register. This means that all the bits at the intersection of the recently failed column and all the previously failed rows are flipped.
 \item At the end of the fourth half iteration, the number of failed rows and columns is checked. 
 \begin{itemize}
  \item If the number of failed rows is zero or more than four, post processing was not applied, and no post-processing iteration is issued.
  \item If the number of failed rows is between one and three, but the number of failed columns is not, post processing was indeed applied, but additional iterations would be useless. In fact, either there are no failed columns (general successful decoding) or there are more than three (the stall pattern is too large and bit flipping will not correct it).
  \item If both row and column failures are between one and three, post processing was applied, and we can hope that we are now out of the stall pattern. A post-processing iteration is issued.
 \end{itemize}
\end{enumerate}

The modified schedule allows the bit flipping step to be performed concurrently with the fourth half iteration, and its performance is  equivalent to the schedule described in \cite{Condo_GlobalSip16}. 

\subsection{Post-Processing Iteration}
\label{subsec:PPit}


The post-processing iteration is issued under the conditions portrayed in Section \ref{subsubsec:PPA}, and it involves up to three rows and three columns. 
During the second iteration, each component decoder stores the indices of the first three failed row (column) decodings. These indices are gathered by the \texttt{control} module that, in case the conditions for a post-processing iteration apply, generates the appropriate control signals (input and output row, output validation) for the row (columns) that were involved in the post-processing application.
To reduce the complexity of the control logic, each post-processing half iteration is always supposed to involve three rows (columns), each decoded in a different clock cycle. Thus, each post-processing half iteration lasts $3+6$ clock cycles, where 6 is the internal component-decoder pipeline depth.

%

\section{Implementation Results and Comparison}
\label{sec:res}

The decoder architecture described in the previous section has been synthesized in TSMC CMOS 65 nm technology using Cadence RTL Compiler, was verified with Mentor Graphics ModelSim and tested with an Altera FPGA. Table~\ref{tab:asic} reports the synthesis results for three target frequencies, in terms of area occupation, gate count, latency and information throughput. The timing constraints have been met for all three frequencies, showing that the proposed architecture can be clocked at 609~MHz, and thus achieve 100~Gb/s of information throughput, even with an older technology node like the 65~nm one. The 193 clock-cycles maximum latency is consistently kept under 1 $\mu$s with all frequencies, while the gate count ranges from 898~kgates at 300~MHz to 1155 kgates when targeting the highest frequency. Supposing that the post processing is applied every time, the design yields a worst case information throughput of 164 bits/cycle. However, post processing is not always necessary, and the post-processing iteration is often not performed. Thus, at a very low BER such as $10^{-15}$, the average throughput tends to the maximum achievable throughput of $181$ bits/cycle.

Very few detailed reports of decoder implementations for OTN hard-decision FEC schemes can be found in the literature. To the best of our knowledge, \cite{Lee_ISOCC10} is the most recent: the considered FEC scheme uses a modified product-like concatenation of long BCH codes, resulting in a code length of almost 4 million bits and a code rate of 0.933. At a BER of $10^{-15}$, \cite{Lee_ISOCC10} has an NCG of 9.19 dB, against the 9.52 dB gained by our scheme (see Table \ref{tab:NCG}). It achieves a throughput of 110~Gb/s with a latency of 38~$\mu$s, while our decoder reaches 100~Gb/s with a 319~ns latency.
The decoder in \cite{Lee_ISOCC10} has been synthesized in 90 nm CMOS technology, and yields a gate count of 3732~kgates at 430 MHz, not including SRAM, against the 1155~kgates of the decoder proposed in this work. Moreover, our decoder only uses registers, no SRAM, and the area these registers occupy is included in the gate count. By comparison, the decoder proposed in \cite{Lee_ISOCC10} utilizes 4 Mbit of SRAM memory.

The more recent braided FEC scheme of \cite{Jian_GLOBECOM13} yields a 9.35 dB NCG at a BER of $10^{-15}$. However, no decoder implementation results were provided. The FEC code length is of 130~kbits with a code rate of 0.937. The decoding process uses a sliding window approach that can limit the gate count, but can have heavy memory requirements while greatly increasing the latency. The latency is estimated at 1.15 million bits.

Soft-decision FECs for OTN have been considered only in recent years: thus, no decoder implementations were found in literature. Considering the gate count and NCG estimations for soft-decision FECs in \cite{Huawei_SDFEC}, it can be seen that the NCG achieved in this work sits in the middle between literature's hard-decision FECs and soft-decision FECs, while the proposed decoder implementation requires an order of magnitude less gates than soft-decision decoders. 



\subsection{FPGA Test and Verification}
\label{subsec:HWtest}

After post-synthesis functional verification with ModelSim, the product decoder has been implemented on an FPGA within a partial digital communication chain. While random data were generated and encoded on a computer, the remainder of the chain has been synthesized to be run on an Altera DE4 board, a board featuring a large Altera Stratix IV EP4SGX530KH40 FPGA. The product decoder easily fits on this FPGA, and enough spare logic is present for the remainder of the communication chain.

\begin{figure}[tbp!]
	\begin{center}
	\includegraphics[scale=0.56]{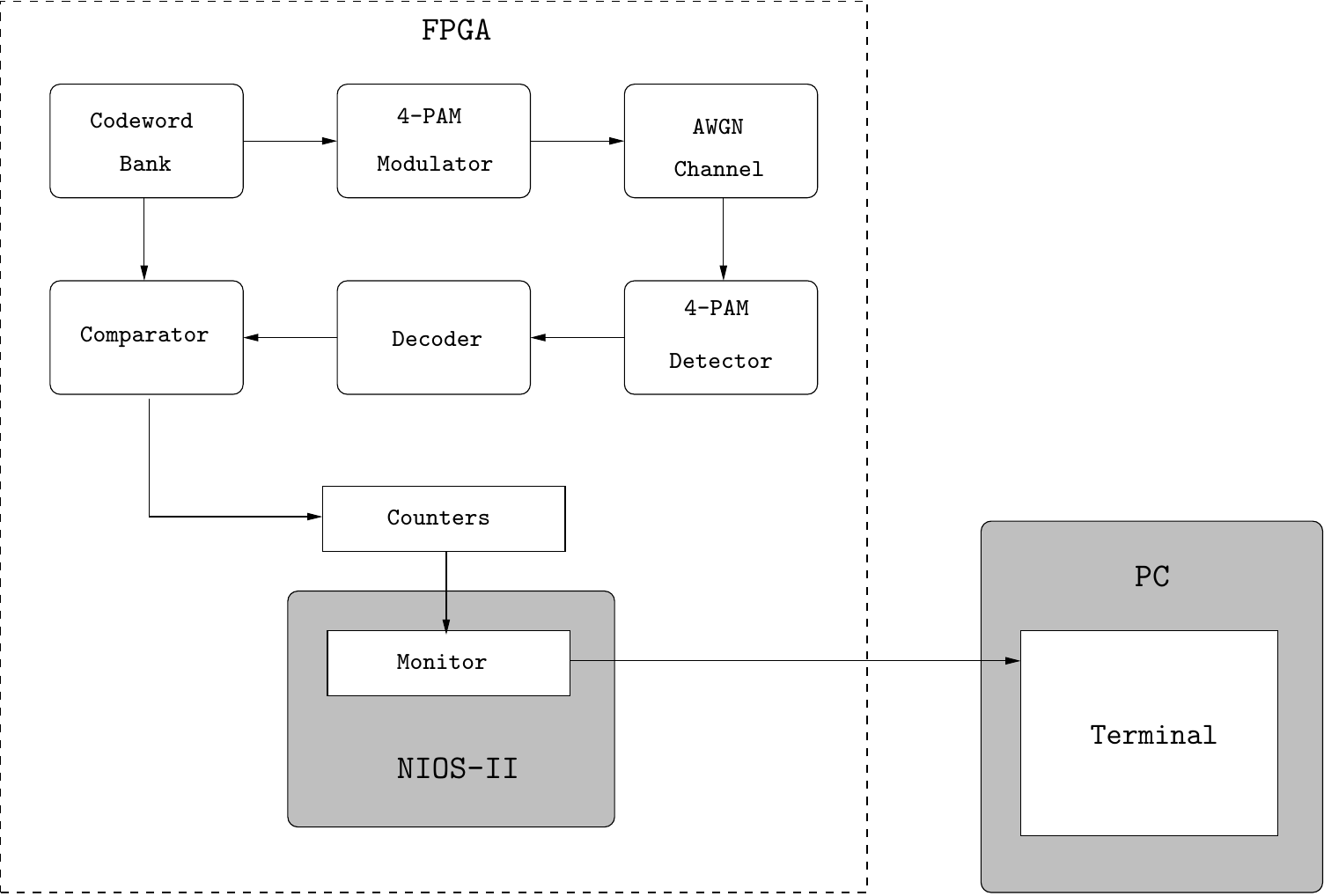}
	\caption{Test methodology with the Altera DE4 board.}
	\label{fig:FPGAimpl}
	\end{center}
\end{figure}

Fig.~\ref{fig:FPGAimpl} shows the experimental setup used for testing. 
The codeword bank stores a set of encoded noiseless codewords. Unlike the software simulations used in the design of the FEC scheme, we considered an Additive White Gaussian Noise (AWGN) channel and 2-bit Pulse-Amplitude Modulation (4-PAM).
The test setup leverages the Nios~II soft-core processor and the UART serial interface over JTAG over USB. As shown in the figure, most of the system is run with dedicated hardware blocks and the software application running on the Nios~II processor is exclusively used to monitor the on-going testing results. Once it has setup the chain, the software application periodically reads the performance counters, calculates $p$ and BER, and pushes the results over the UART-over-JTAG-over-USB link to a terminal running on the host PC.

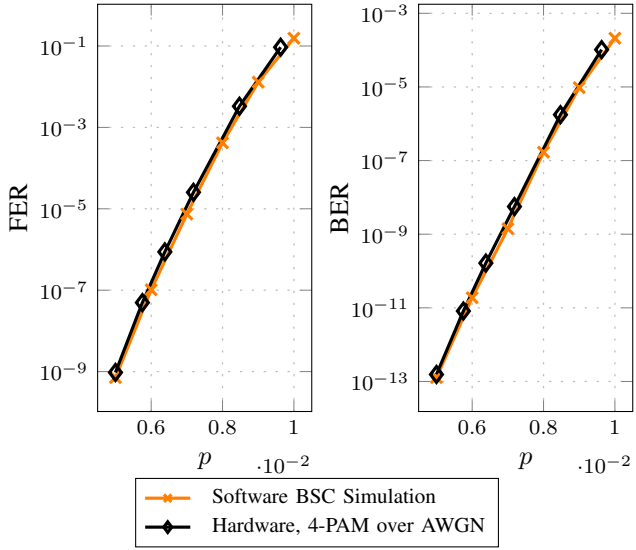
\begin{figure}[tbp!]
	\begin{center}




\begin{tikzpicture}

  \pgfplotsset{
    grid style = {
      dash pattern = on 0.05mm off 1mm,
      line cap = round,
      line width = 0.5pt,
      loosely dotted
    },
    label style = {font=\fontsize{10pt}{7.2}\selectfont},
    tick label style = {font=\fontsize{8pt}{7.2}\selectfont}
  }

  \begin{semilogyaxis}[%
    xlabel=$p$,%
    ylabel=FER, ylabel style={yshift=-0.5em},%
    every axis x label/.style={at={(ticklabel cs:0.5)},anchor=near ticklabel},
    width=0.5\columnwidth, height=7cm, grid=major,%
    legend style={
      anchor={center},
      cells={anchor=west},
      column sep=2mm,
      font=\fontsize{8pt}{7.2}\selectfont,
    },
    legend to name=sim-vs-hw-legend,
    mark size=3.0pt,
    mark options=solid,
    grid=major,
    xminorgrids=true,
    yminorgrids=true,
    ]

    \addplot[color=orange, mark=x, very thick] table[x=p,y=FER]{Figures/data/tpc_195-178-2_2iter_pp1_cd3f357.txt};
    \addlegendentry{Software BSC Simulation}
    \addplot[color=black,mark=diamond, very thick] table[x=p,y=FER]{Figures/data/tpc_195-178-2_2iter_pp1_hw-unlimited.txt};
    \addlegendentry{Hardware, 4-PAM over AWGN}
  \end{semilogyaxis}
\end{tikzpicture}
\begin{tikzpicture}
  \pgfplotsset{
    grid style = {
      dash pattern = on 0.05mm off 1mm,
      line cap = round,
      line width = 0.5pt,
      loosely dotted
    },
    label style = {font=\fontsize{10pt}{7.2}\selectfont},
    tick label style = {font=\fontsize{8pt}{7.2}\selectfont}
  }
  \begin{semilogyaxis}[%
    xlabel=$p$,%
    ylabel=BER, ylabel style={yshift=-0.5em},%
    every axis x label/.style={at={(ticklabel cs:0.5)},anchor=near ticklabel},
    width=0.5\columnwidth, height=7cm, grid=major,%
    mark size=3.0pt,
    mark options=solid,
    grid=major,
    xminorgrids=true,
    yminorgrids=true,
    ]

    \addplot[color=orange, mark=x, very thick] table[x=p,y=BER]{Figures/data/tpc_195-178-2_2iter_pp1_cd3f357.txt};
    \addplot[color=black, mark=diamond, very thick] table[x=p,y=BER]{Figures/data/tpc_195-178-2_2iter_pp1_hw-unlimited.txt};
  \end{semilogyaxis}
\end{tikzpicture}
\\
\ref{sim-vs-hw-legend}

  \caption{Error-correction performance comparison between software simulation and hardware results.}
  \label{fig:sim_vs_hw}
	\end{center}
\end{figure}

Clocked at 50 MHz, the test setup shows an average measured information throughput of 9.98 Gb/s in the regions of interest, equivalent to a coded throughput of 11.98 Gb/s. Fig.~\ref{fig:sim_vs_hw} shows a comparison of the expected error-correction performance---frame-error rate (FER) on the left and BER on the right---compared to that of the hardware implementation. Software simulations are for a BSC. For the hardware implementation a bank of 64 random codewords generated with the software encoder are modulated on a Gray-coded 4-PAM constellation. With a slight abuse of notation, we refer to the decoder's input BER with the AWGN channel as $p$ as well. The AWGN channel has been simulated through an open source Gaussian noise generator available on OpenCores.org \cite{gngwebsite}. A 4-PAM detector finally generates the hard values that are fed to the decoder.

In hardware, the communication chain was run until a minimum of $1 \times 10^{8}$ frames were decoded and at least 100 frames were found to be in error. As both conditions were required, the last point of the hardware curves translated into the decoding of over $1 \times 10^{11}$ frames.

From Fig.~\ref{fig:sim_vs_hw}, it can be seen that the hardware and software simulation curves---solid black and orange with diamond markers, respectively---are very close to each other. The small differences are likely attributed to the different channels, and the use of a fixed-point number representation for both modulation and noise versus a floating-point one in software. Furthermore, the decoder implementation alone was simulated at the RTL level to be bit true with the software model for thousands of frames.


\begin{table}[tbp!]
  \centering
    \caption{TSMC CMOS 65 nm ASIC synthesis results. }
    		\setlength{\extrarowheight}{1.8pt}
  \begin{tabular}{|c|c|c|c|}

  
    \hline
    Target Frequency [MHz]& 300 &400 & 609\\ 
    Area [mm$^2$]& 1.052 & 1.134 & 1.352 \\ 
    Gate Count [kgates]& 898 &968 &1155\\ 
    Latency [ns]& 643 & 483 & 319\\ 
    $T$ [Gb/s] &49.2 & 65.7 & 100\\ 
    $T$ [bits/cycle] &164 & 164 & 164\\ 
    \hline
	\end{tabular}

  \label{tab:asic}
\end{table}

\section{Architectural Modifications}
\label{sec:mods}

In this Section, we briefly consider possible modifications to the decoder architecture in case of changes to the code parameter or to the specified constraints.

The product decoder is completely rate flexible: as long as the code length remains $195^2$, no modifications are required if the number of information bits becomes something else than $178^2$.

Increasing or decreasing the number of performed standard iterations is a straightforward modification of two configuration parameters. It requires that the maximum value of the iteration counter be changed, along with the iteration value at which post processing is applied.

A change of $t$ requires a different decoding algorithm, so the decoder must be completely redesigned.

A change in code length (meaning a different shortening value, but with the same root BCH code) mandates radical changes to all modules of the decoder. It will affect the size of the scratch memory, the number of rows/columns connected to each component decoder, and the structure of the component decoders themselves. While it is true that the decoding algorithm remains the same, since $t$ is not changed, most eBCH decoder modules are code-length specific and fine-tuned to the proposed FEC scheme. The \texttt{Selectors and Logarithms} and \texttt{Error Locator} modules require minor modifications to accommodate the longer code, but \texttt{Parity} and \texttt{Syndrome} multilevel XOR trees must be redesigned, and similarly the bit flipping signal generation algorithm implemented in the \texttt{Bit flipping and post-processing} module.



The proposed decoder architecture relies on $P_c=13$ component decoders, that are able to achieve the $100$ Gb/s information-throughput specifications with a clock frequency of $609$~MHz.
The number of clock cycles required to decode a $(195,178)^2$ product codeword can be expressed as follows:
\begin{flalign}\label{eq:time}
  &\left(\bigg\lfloor \frac{P_c}{P_l}\bigg\rfloor + \min\bigg(P_c - P_l\bigg\lfloor \frac{P_c}{P_l}\bigg\rfloor, P_l  \bigg)\right)\cdot \bigg\lfloor \frac{195}{P_c} \bigg\rfloor + \\
  ~+ \bigg( 2&02 - P_c \bigg\lfloor \frac{195}{P_c} \bigg\rfloor\bigg) +  (2L-1)\bigg\lceil\frac{195}{P_c} \bigg\rceil + (2+2L)n_p &&\nonumber
\end{flalign}
where $P_l$ is the number of 195-bit loading lanes (currently $2$), $n_p$ is the number of pipeline stages in the component decoders (currently 6) and $L$ the number of decoding iterations excluding the post-processing iteration (currently 2). Consequently, the decoding process amounts to 193 clock cycles or:
\begin{itemize}
 \item 111 clock cycles for the loading of the codeword and the concurrent execution of the first half-iteration;
 \item 21 clock cycles for the following three half-iterations, for a total of 63 clock cycles;
 \item 18 clock cycles for the post-processing iteration;
 \item 1 clock cycle to signal the end of the decoding.
 \end{itemize}

In case throughput requirements are lower, or in case the achievable frequency is higher than $609$ MHz, the decoder can be redesigned to meet the new specifications.
For example, if the decoder was to be implemented with a deep sub-micron technology node, e.g. CMOS 28 nm, an achievable clock frequency of 1~GHz would likely be possible. In this situation, the $100$ Gb/s information-throughput constraint would be met whenever the decoding process lasts at most 316 clock cycles.
In this case, a higher number of iterations $L$ or a lower number of component decoders $P_c$ might be considered.

\section{Conclusions}
\label{sec:conc}

In this work, we have proposed a novel FEC scheme for OTN. It uses product codes with extended-BCH codes as component codes and a post-processing technique that greatly reduces the error floor. The proposed FEC achieves 9.52~dB of NCG at a BER of $10^{-15}$ and 9.96 dB at $10^{-18}$. A low-complexity, high-speed decoder architecture has been designed, tested on FPGA and synthesized in 65 nm CMOS technology: it yields a worst-case throughput of 164 bits/cycle, i.e. an information throughput of 100 Gb/s at 609~MHz, with a gate count of 1.15 million gates. The proposed FEC brings the error-correction performance of hard-decision FECs closer to that of soft-decision FECs. The complexity of the proposed decoder is lower than that of hard-decision decoders in literature, and an order of magnitude lower than the estimated complexity of soft-decision decoders. The $319$~ns latency makes the proposed FEC scheme and decoder suitable for low-latency environments like data centers.

\bibliographystyle{IEEEtran}
\bibliography{IEEEabrv,refs}

\begin{thebibliography}{10}
\providecommand{\url}[1]{#1}
\csname url@samestyle\endcsname
\providecommand{\newblock}{\relax}
\providecommand{\bibinfo}[2]{#2}
\providecommand{\BIBentrySTDinterwordspacing}{\spaceskip=0pt\relax}
\providecommand{\BIBentryALTinterwordstretchfactor}{4}
\providecommand{\BIBentryALTinterwordspacing}{\spaceskip=\fontdimen2\font plus
\BIBentryALTinterwordstretchfactor\fontdimen3\font minus
  \fontdimen4\font\relax}
\providecommand{\BIBforeignlanguage}[2]{{%
\expandafter\ifx\csname l@#1\endcsname\relax
\typeout{** WARNING: IEEEtran.bst: No hyphenation pattern has been}%
\typeout{** loaded for the language `#1'. Using the pattern for}%
\typeout{** the default language instead.}%
\else
\language=\csname l@#1\endcsname
\fi
#2}}
\providecommand{\BIBdecl}{\relax}
\BIBdecl

\bibitem{Lee_ISOCC10}
K.~Lee and H.~Lee, ``A high-performance concatenated {BCH} code and its
  hardware architecture for 100 {Gb/s} long-haul optical communications,'' in
  \emph{Int. SoC Design Conf. (ISOCC)}, Nov 2010, pp. 428--431.

\bibitem{Smith_JLT12}
B.~P. Smith, A.~Farhood, A.~Hunt, F.~R. Kschischang, and J.~Lodge, ``Staircase
  codes: {FEC} for 100 {G}b/s {OTN},'' \emph{J. Lightw. Technol.}, vol.~30,
  no.~1, pp. 110--117, Jan 2012.

\bibitem{Jian_GLOBECOM13}
Y.-Y. Jian, H.~Pfister, K.~Narayanan, R.~Rao, and R.~Mazahreh, ``Iterative
  hard-decision decoding of braided {BCH} codes for high-speed optical
  communication,'' in \emph{IEEE Global Commun. Conf. (GLOBECOM)}, Dec 2013,
  pp. 2376--2381.

\bibitem{Gallager_TIT62}
R.~Gallager, ``Low-density parity-check codes,'' \emph{IRE Trans. Inf. Theory},
  vol.~8, no.~1, pp. 21--28, January 1962.

\bibitem{Onohara_JSTQE10}
K.~Onohara, T.~Sugihara, Y.~Konishi, Y.~Miyata, T.~Inoue, S.~Kametani,
  K.~Sugihara, K.~Kubo, H.~Yoshida, and T.~Mizuochi, ``Soft-decision-based
  forward error correction for 100 {Gb/s} transport systems,'' \emph{{IEEE} J.
  Sel. Topics Quantum Electron.}, vol.~16, no.~5, pp. 1258--1267, Sept 2010.

\bibitem{Sugihara_OFC13}
K.~Sugihara, Y.~Miyata, T.~Sugihara, K.~Kubo, H.~Yoshida, W.~Matsumoto, and
  T.~Mizuochi, ``A spatially-coupled type {LDPC} code with an {NCG} of 12 {dB}
  for optical transmission beyond 100 {G}b/s,'' in \emph{Opt. Fiber Commun.
  Conf. and Exposition and the Nat. Fiber Opt. Eng. Conf. (OFC/NFOEC)}, March
  2013, pp. 1--3.

\bibitem{Huawei_SDFEC}
\BIBentryALTinterwordspacing
Huawei. Soft-decision {FEC}: Key to high-performance {100G} transmission.
  [Online]. Available:
  \url{www.huawei.com/ilink/en/solutions/broader-smarter/morematerial-b/HW\_112021}
\BIBentrySTDinterwordspacing

\bibitem{Fujitsu_SDFEC}
\BIBentryALTinterwordspacing
Fujitsu. Soft-decision {FEC} benefits for {100G}. [Online]. Available:
  \url{http://www.fujitsu.com/ca/en/Images/Soft-Decision-FEC-Benefits-or-100G-wp.pdf}
\BIBentrySTDinterwordspacing

\bibitem{Bose_IC60}
R.~Bose and D.~Ray-Chaudhuri, ``On a class of error correcting binary group
  codes,'' \emph{Inf. Control}, vol.~3, no.~1, pp. 68 -- 79, 1960.

\bibitem{Wang_CL12}
Z.~Wang, ``Super-{FEC} codes for 40/100 {G}bps networking,'' \emph{{IEEE}
  Commun. Lett.}, vol.~16, no.~12, pp. 2056--2059, Dec 2012.

\bibitem{Miyata_OECC13}
Y.~Miyata, K.~Kubo, K.~Sugihara, T.~Ichikawa, W.~Matsumoto, H.~Yoshida, and
  T.~Mizuochi, ``Performance improvement of a triple-concatenated {FEC} by a
  {UEP-BCH} product code for 100 {G}b/s optical transport networks,'' in
  \emph{OptoElectron. and Commun. Conf. (OECC/PS)}, Jun 2013, pp. 1--2.

\bibitem{Elias_IRE54}
P.~Elias, ``Error-free coding,'' \emph{Trans. IRE Prof. Group Inf. Theory},
  vol.~4, no.~4, pp. 29--37, September 1954.

\bibitem{Gorenstein}
D.~Gorenstein, W.~W. Peterson, and N.~Zierler, ``Two-error correcting
  {Bose-Chaudhuri} codes are quasi-perfect,'' \emph{Inf. Control}, vol.~3,
  no.~3, pp. 291--294, 1960.

\bibitem{Condo_GlobalSip16}
\BIBentryALTinterwordspacing
C.~Condo, F.~Leduc-Primeau, G.~Sarkis, P.~Giard, and W.~J. Gross, ``Stall
  pattern avoidance in polynomial product codes,'' in \emph{{IEEE} Global Conf.
  on Signal and Inf. Process. ({GlobalSIP})}, Dec 2016, to appear. [Online].
  Available: \url{http://arxiv.org/abs/1611.04834}
\BIBentrySTDinterwordspacing

\bibitem{Onohara_OFC10}
K.~Onohara, Y.~Miyata, T.~Sugihara, K.~Kubo, H.~Yoshida, and T.~Mizuochi,
  ``Soft decision {FEC} for {100G} transport systems,'' in \emph{Opt. Fiber
  Commun. Conf. (OFC), collocated Nat. Fiber Opt. Eng. Conf. (OFC/NFOEC)},
  March 2010, pp. 1--3.

\bibitem{gngwebsite}
\BIBentryALTinterwordspacing
G.~Liu. Gaussian noise generator. [Online]. Available:
  \url{http://opencores.org/project,gng}
\BIBentrySTDinterwordspacing

\end{thebibliography}

\end{document}